\documentclass{elsart} %Change as appropriate 
\journal{Phys. Lett. A}

\usepackage[T1]{fontenc} % Font encoding.
\usepackage[latin1]{inputenc} % Input encoding - change for other
\usepackage{amsmath} % Extra environments etc.
\usepackage{amsfonts} % Additional fonts available.
\usepackage{amssymb}
\usepackage[english]{babel} % Multiple languages support.
\usepackage{enumerate} % Better enumerate environment

\usepackage{palatino, mathpple} % Palatino
\newcommand{\norm}[1]{\ensuremath{\left\Vert #1 \right\Vert}}
\newcommand{\abs}[1]{\ensuremath{\left\vert #1 \right\vert}}
\newcommand{\dZ}{\ensuremath{\mathbb{Z}}}
\newcommand{\dR}{\ensuremath{\mathbb{R}}}

\begin{document}

\begin{frontmatter}
\title{Diophantine approximation and the solubility of the
  Schrödinger equation}
\author{Simon Kristensen\thanksref{thanks}}
\thanks[thanks]{Research funded by EPSRC grant no. GR/N02832/01}
\address{Department of Mathematics \\
  University of York \\
  Heslington \\
  York, YO10 5DD \\
  United Kingdom \\
  Tel.: (+44) [0]1904 433072 \\
  Fax.: (+44) [0]1904 433071 \\
}
\ead{sk17@york.ac.uk}

\begin{abstract}
  We characterise the set of periods for which number theoretical
  obstructions prevent us from finding periodic solutions of the
  Schrödinger equation on a two dimensional torus as well as the
  asymptotic occurrence of possible resonances.
\end{abstract}

\begin{keyword}
The Schrödinger equation \sep Diophantine approximation \sep
Hausdorff dimension 
\def\MSC{\par\leavevmode\hbox {\it 2000 MSC:\ }}
\PACS 03.65.-w \sep 03.65.Db
\MSC 81Q05 \sep 11K60
\end{keyword}

\end{frontmatter}

\section{Introduction}
\label{sec:introduction}

Number theoretical considerations arise naturally in the theory of 
partial differential equations in connection with the notorious
problem of small denominators. In this paper, we examine such a
problem occurring in the Schrödinger equation for a particle moving on
a two-dimensional torus in a potential. We also briefly discuss the
corresponding but much more difficult problem associated with the
classical wave equation.

A natural way to study the solubility of partial differential
equations on a torus is using Fourier series. Solving these equations
for the Fourier coefficients formally yields ratios in which the
denominators are on the form  
\begin{displaymath}
  a_1 x_1 + \cdots + a_k x_k + a_0,
\end{displaymath}
where the $x_j$ are dependent on the partial differential equation in
question and the $a_i$ are drawn from certain subsets of the integers,
with each choice of $a_i$'s corresponding to an eigenfrequency. \sloppy{When
these denominators become small, the corresponding eigenvalue
may become large, depending on the value of the numerator.} If too many
of the denominators become too small, the Fourier series may fail to
converge. Diophantine approximation is a suitable tool for studying
the exceptional set in which this happens.

The techniques needed for the resolution of the small denominator
problem for the Schrödinger equation on the two-dimensional torus have
essentially been obtained in a purely number theoretical setting
\cite{MR93a:11066,MR2001g:11129,MR28:3018}. However, those papers made
no connection between the number theoretical results and physics. In
this paper, the connection between the theory of Diophantine
approximation and the Schrödinger equation is made, and the relevant
results are adapted to the particular small denominator problem
associated with the Schrödinger equation.

\section{Solubility of the Schrödinger equation}
\label{sec:solub-schr-equat}

For a single particle on a two dimensional torus with dimensions
$\alpha \times \beta$, the Schrödinger equation reads
\begin{equation}
  \label{eq:1}
  i \hslash \frac{\partial u(x,y,t)}{\partial t} +
  \frac{\hslash^2}{2m} \nabla^2 u(x,y,t) = V(x,y,t)u(x,y,t),
\end{equation}
where $V$ denotes a smooth potential on the torus. We assume that the
potential is periodic in $t$ with period $\gamma > 0$. We are looking
for smooth solutions $u$ with the same period in $t$.

We may interpret the above setup as considering a particle moving
according to the Schrödinger equation in the compact spacetime given
by the three dimensional torus with dimensions $\alpha \times \beta
\times \gamma$ and with signature $(+,+,-)$. This spacetime is
acausal, as it contains closed timelike curves. Thus, in discussing
the solubility of the Schrödinger equation in this setting, we are in
fact examining whether or not such a spacetime is consistent with
quantum physics and in particular with the Schrödinger equation. In
popular terms, we are examining whether or not quantum physics objects
to certain types of time machines.

As we are assuming that both $u$ and $V$ are periodic in all three
variables, we may expand them in their Fourier series,
\begin{equation}
  \label{eq:5}
  u(x,y,t) = \sum_{a,b,c \in \dZ} u_{a,b,c} \exp\left(2 \pi i
    \left[\frac{a}{\alpha}x + \frac{b}{\beta}y +
      \frac{c}{\gamma}t\right]\right)
\end{equation}
and \begin{displaymath} V(x,y,t) = \sum_{a,b,c \in \dZ}
  V_{a,b,c} \exp\left(2 \pi i \left[\frac{a}{\alpha}x +
      \frac{b}{\beta}y + \frac{c}{\gamma}t\right]\right).
\end{displaymath}
Assume for the moment that we have a solution $u$ to \eqref{eq:1} and
define the forcing term
\begin{displaymath}
  f(x,y,t) = u(x,y,t)V(x,y,t) =  \sum_{a,b,c \in \dZ}
  f_{a,b,c} \exp\left(2 \pi i \left[\frac{a}{\alpha}x +
      \frac{b}{\beta}y + \frac{c}{\gamma}t\right]\right)
\end{displaymath}
where 
\begin{equation}
  \label{eq:8}
  f_{a,b,c} = \sum_{a',b',c' \in \dZ} u_{a',b',c'}
  V_{a-a',b-b',c-c'}. 
\end{equation}
Now, equation \eqref{eq:1} becomes
\begin{equation}
  \label{eq:9}
  i \hslash \frac{\partial u(x,y,t)}{\partial t} +
  \frac{\hslash^2}{2m} \nabla^2 u(x,y,t) = f(x,y,t). 
\end{equation}
Consider the corresponding homogeneous equation where $f = 0$. In this
case, inserting the Fourier series for $u$ in \eqref{eq:9} yields the
only solution $u = 0$. 

Integrating on both sides and inserting the Fourier series, we see
that $\iiint f dx dy dt = 0$ or equivalently $f_{0,0,0} = 0$. This
follows by the periodicity requirements and the fundamental theorem of
calculus. This in turn implies that for the solution $u$, we must have
$u_{0,0,0} = 0$.

To obtain the required Fourier coefficients for $u$ when $(a,b,c) \neq
(0,0,0)$, we insert the Fourier series for $u$ and $f$ in \eqref{eq:9}
and identify the coefficients to obtain a formula for each of the
Fourier coefficients of $u$ in terms of the coefficients for the
forcing term $f$,
\begin{equation}
  \label{eq:2}
  u_{a,b,c} = - \frac{\gamma}{2 \pi \hslash}
      \frac{f_{a,b,c}}{\frac{\pi \hslash \gamma}{m \alpha^2} a^2
      +\frac{\pi \hslash \gamma}{m \beta^2}b^2 + c}.
\end{equation}

We are assuming that a smooth solution does indeed exist, so that
\eqref{eq:2} makes sense. This gives rise to two sufficient conditions
for a solution to \eqref{eq:1} to exist. The first condition relates
to the numerator in \eqref{eq:2}. Since both $V$ and $u$ are assumed
to be smooth, the Fourier coefficients of the product must decay
faster than the reciprocal of any polynomial. This implies that for
any polynomial $P$,  
\begin{equation} \tag{C1}
  \label{cond:1} 
  \abs{f_{a,b,c}} = \abs{\sum_{a',b',c' \in \dZ} u_{a',b',c'}
    V_{a-a',b-b',c-c'}} = o\left(P(a,b,c)^{-1}\right).
\end{equation}
where we also require that 
\begin{equation} \tag{C1'}
  \label{cond:1'}
  f_{0,0,0} = \sum_{a',b',c' \in \dZ} u_{a',b',c'} V_{-a',-b',-c'}
  = 0. 
\end{equation}
Condition \eqref{cond:1'} is equivalent to the requirement that
$\iiint f dx dy dt= 0$. These condition relates the wave function to
the potential.

The second condition relates to the denominator in \eqref{eq:2}. The
Fourier expansion of $u$ will converge if we have for some $v > 0$ and
some $K > 0$,
\begin{equation}
  \label{cond:2} \tag{C2}
  \abs{\dfrac{\pi \hslash \gamma}{2 m \alpha^2} a^2 +\dfrac{\pi
      \hslash \gamma}{2 m \beta^2}b^2 + c} > K \max(a^2, b^2)^{-v} 
  \text{ for any } a,b,c \in \dZ.
\end{equation}
This condition relates the possible periods in $t$ to the possible
dimensions of the tori for which the Schrödinger equation \eqref{eq:1}
may be solved. If conditions \eqref{cond:1}, \eqref{cond:1'} and
\eqref{cond:2} are satisfied, \eqref{eq:5} defines a smooth function
on the torus, which satisfies \eqref{eq:1}. Thus these are indeed
sufficient for a solution to exist.

A word about the periodicity requirement on $u$. Inspired by Floquet's
Theorem, it is reasonable to conjecture that solutions of the form
$e^{\lambda t} u(x,y,t)$ would exist with $u$ periodic. However, on
substituting this expression into \eqref{eq:1}, dividing by
$e^{\lambda t}$ and rearranging the terms gives us an equation of the
same form as \eqref{eq:1} with $V(x,y,t)$ replaced by $V(x,y,t) - i
\hslash \lambda$ which $u$ must satisfy in order for the more general
function to satisfy the original equation. Thus, the same obstructions
exist when we consider solutions of this form. Note also, that for
such solutions the topology of the spacetime changes, and we are no
longer considering the same physical system. Even so, in this
non-compact spacetime, the problem of small denominators persists.

In this paper, we are interested in the small denominator problem
associated with the failure of Condition \eqref{cond:2} to hold. This
means that in effect, we are studying the partial differential
equation given by \eqref{eq:9}.  where $f$ is smooth with $\iiint f =
0$. Conversely, all our results apply to equations of that type.

For any $x \in \dR$, we let $\norm{x}$ denote the distance from $x$
to the nearest integer. Let $v > 0$. We define the set
\begin{multline}
  \label{eq:3}
  \mathcal{E} = \bigg\{(x,y) \in [0,1]^2 : \norm{a^2 x + b^2
  y} < \max\left(a^2, b^2\right)^{-v}\\ \text{ for infinitely many }
  (a,b) \in \dZ^2\bigg\}.
\end{multline}
For any $(\tfrac{\pi \hslash \gamma}{m \alpha^2}, \tfrac{\pi \hslash
  \gamma}{m \beta ^2}) \in \mathcal{E}$, Condition \eqref{cond:2}
fails to hold.

More general forms of the set $\mathcal{E}$ have been studied
by Schmidt \cite{MR28:3018} and Rynne \cite{MR93a:11066}. The latter
result was generalised to even more general sets by Dickinson and
Rynne \cite{MR2001g:11129}. Their results relating to the set
$\mathcal{E}$ are summarised in the following theorem.
\begin{thm}[Corollary to \cite{MR28:3018} and \cite{MR93a:11066}]
  \label{thm:main}
  Let $v > 0$. The Lebesgue measure of the set $\mathcal{E}$
  is full when $v \leq 1$ and null when $v > 1$. When $ v > 1$, the
  Hausdorff dimension of $\mathcal{E}$ is $ 1 + 2/(v+1)$.

  Furthermore, if we let $\epsilon > 0$ and $N(k,v;x,y)$ denote the
  number of solutions to
  \begin{displaymath}
    \norm{a^2 x + b^2 y} < \max\left(a^2, b^2\right)^{-v}
  \end{displaymath}
  with $1 \leq a,b \leq k$, then for almost all $x,y \in \dR$,
  \begin{multline}
    \label{eq:4}
    N(k,v;x,y) = \left(\sum_{h=1}^k \dfrac{1}{h^v}\right)^2 +
    O\left(\sum_{h=1}^k \dfrac{1}{h^v}\right)^{1+\epsilon} \\ =
    \begin{cases}
      O(1) & \text{when } v > 1 \\ (\log k)^2 + O \left(\log
      k\right)^{1+\epsilon} & \text{when } v = 1 \\
      \tfrac{1}{v+1}k^{2(v+1)} +
      O\left(\tfrac{1}{v+1}k^{(v+1)}\right)^{1+\epsilon} & \text{when
      } v < 1.
    \end{cases}
  \end{multline}
\end{thm}

Note that the final asymptotic formula implies the measure results of
the theorem. The theorem characterises the values of $(\tfrac{\pi
\hslash \gamma}{m \alpha^2}, \tfrac{\pi \hslash \gamma}{m \beta
^2})$ for which Condition \eqref{cond:2} fails to hold. Indeed, the
set of such points must have measure zero and even Hausdorff dimension
$1$, as the condition is required to hold for all $v >
0$. 

In physical terms, The first part of this result states that for most
of the spacetimes discussed, the Schrödinger equation may be solved
using Fourier series. Furthermore, \eqref{eq:4} characterises the
number of small denominators occurring for $1 \leq a,b \leq k$. Each
such small denominator causes the Fourier coefficient to become large,
and thus corresponds to a comparatively large frequency in the
spectrum of the wave function.

\begin{rem}
  A direct but rather lengthy proof of the measure and dimension
  results of Theorem \ref{thm:main} is possible, using methods from
  Dodson's paper \cite{MR95d:11092}. 
\end{rem}

A related problem is the problem of characterising the problematic
periods in the classical, inhomogeneous wave equation,
\begin{displaymath}
  \dfrac{\partial^2 u (t,x,y)}{\partial t ^2} - \nabla^2 u (t,x,y) =
  f(t,x,y).
\end{displaymath}
On the one dimensional torus, this was studied by Nov{\'a}k
\cite{MR50:7828} and later in more generality by Fe{\v{c}}kan
\cite{MR95c:35030} and in even higher generality by Gramchev and
Yoshino \cite{MR96m:35048}, who obtained necessary and sufficient
conditions for solubility of a large class of partial differential
equations including the wave equation. However, the problem is
unsolved for the two dimensional torus. In this case, the analogue of
Condition \eqref{cond:1} disappears, as $u$ does not occur on the
right hand side. Condition \eqref{cond:1'} reduces to the condition
that $\iiint f = 0$. The Diophantine condition corresponding to
Condition \eqref{cond:2} becomes
\begin{equation}
  \label{eq:6}
  \abs{a^2 x + b^2 y - c^2} > K \max(a^2,b^2)^{-v}
\end{equation}
for some $v > 0$ and some $K>0$ for all $(a,b,c) \in \mathbb{Z}^3$
with $(a,b) \neq (0,0)$. From Theorem \ref{thm:main}, it follows that
the set for which this fails to be the case has measure zero and
dimension $1$. However, finding the critical exponent $v$ where the
measure of the set associated with the failure of \eqref{eq:6} drops
from full to null and finding the dimension of this set in the
null-region turns out to be very difficult.

It should be noted that since
\begin{displaymath}
  a^2 x + b^2 y - c^2 = \left(\sqrt{a^2 x + b^2 y} - \abs{c}\right)
  \left(\sqrt{a^2 x + b^2 y} + \abs{c}\right),
\end{displaymath}
we may instead of \eqref{eq:6} consider the condition 
\begin{equation}
  \label{eq:7}
  \abs{\sqrt{a^2 x + b^2 y} - \abs{c}} > K' \max(a^2,b^2)^{-v'}
\end{equation}
for some $v' > 0$ and some $K'>0$ for all $(a,b,c) \in \mathbb{Z}^3$
with $(a,b) \neq (0,0)$. This follows as the second term in the
product is always positive (in fact it is to the order of
$\max(\abs{a},\abs{b})$). Unfortunately, analysing the Diophantine
condition \eqref{eq:7} appears to be no less difficult than the
original problem \eqref{eq:6}. 

\section*{Acknowledgements}
I thank Todor V. Gramchev for his inspirational comments and insight.
I also thank Chris Fewster for pointing out the significance of closed
timelike curves and Maurice Dodson for his comments on the various
drafts of the paper. Finally, I thank the referees for their helpful
comments.

\end{document}